\documentclass[prb, aps, twocolumn, superscriptaddress]{revtex4}
\usepackage{graphicx}% Include figure files
\usepackage{bm}% bold math
\usepackage{amssymb}
\usepackage{amsmath}
\usepackage{amsfonts}
\usepackage{array}

\newcommand {\bra} [1] {\langle #1 |}
\newcommand {\ket} [1] {| #1 \rangle}
\newcommand {\bkt} [1] {\langle #1 \rangle}

\newcommand {\pd} [2] {\frac{\partial #1}{\partial #2}}
\newcommand {\td} [2] {\frac{d #1}{d #2}}

\newcommand* {\kk}{{\bm{k}}}
\DeclareMathOperator{\trace}{tr}

\begin{document}
\title{Side-jumps in the spin-Hall effect: construction of the
 Boltzmann collision integral}

\author{Dimitrie Culcer}
\affiliation{Condensed Matter Theory Center, Department of Physics,
 University of Maryland, College Park MD20742-4111}
\author{E. M. Hankiewicz}
\affiliation{Institut f\"ur Theoretische Physik und Astrophysik,
Universit\"at W\"urzburg, 97074 W\"urzburg, Germany}
\author{Giovanni Vignale}
\affiliation{Department of Physics and Astronomy, University of
 Missouri, Columbia, Missouri 65211}
\author{R. Winkler}
\affiliation{Materials Science Division, Argonne National
Laboratory, Argonne, IL 60439}
\affiliation{Northern Illinois University, De Kalb, Illinois 60115}

\begin{abstract}
  We present a systematic derivation of the side-jump contribution
  to the spin-Hall current in systems without band structure
  spin-orbit interactions, focusing on the construction of the
  collision integral for the Boltzmann equation. Starting from the
  quantum Liouville equation for the density operator we derive an
  equation describing the dynamics of the density matrix in the
  first Born approximation and to first order in the driving
  electric field. Elastic scattering requires conservation of the
  total energy, including the spin-orbit interaction energy with the
  electric field: this results in a first correction to the
  customary collision integral found in the Born approximation. A
  second correction is due to the change in the carrier position
  during collisions. It stems from the part of the density matrix
  off-diagonal in wave vector. The two corrections to the collision
  integral add up and are responsible for the total side-jump
  contribution to the spin-Hall current. The spin-orbit-induced
  correction to the velocity operator also contains terms diagonal
  and off-diagonal in momentum space, which together involve the
  total force acting on the system. This force is explicitly shown
  to vanish (on the average) in the steady state: thus the total
  contribution to the spin-Hall current due to the additional terms
  in the velocity operator is zero.
\end{abstract}
\date{\today}
\maketitle

\section{Introduction}

Semiconductor spin electronics is an active area of research in
which both theory and experiment have made substantial progress
during the past decade. \cite{Zutic04, Awschalom07} The
recent focus on electrically-induced phenomena in systems with
spin-orbit interactions has brought to light many unexplored and
fascinating facets of semiconductor physics. Considerable progress
has been made in past years in the electrical manipulation of spins
in semiconductors, where experimental and theoretical work have
yielded the prediction \cite{Dyakonov71, Hirsch99, Zhang00,
 Murakami03, Sinova04, Handbook, WinRev} and discovery of the
spin-Hall effect. \cite{Kato04, Sih05, Wunderlich05, Stern08} The
spin-Hall effect, \cite{Zutic04, Handbook} which is the main focus
of this paper, is the generation of a transverse spin-current
\cite{Culcer07, WinRev} at the edges of the sample as a response to
an external electric field. Such a spin current leads to a
spin-accumulation at the edges of the sample, and the relationship
between spin currents and spin accumulation is nontrivial.
\cite{TsePRB05, GalitskiPRB06} The first observations of the spin-Hall effect
were followed by the measurement of the spin-Hall effect at room
temperature by optical techniques \cite{Stern06} and the first
successful measurement of the spin-Hall effect in transport in
ballistic HgTe/HgCdTe quantum wells. \cite{Brune08}

Two main mechanisms have been shown to be responsible for the
spin-Hall effect. The presence of spin-orbit coupling in the
impurity potentials gives rise to contributions to the spin-Hall
effect which are termed \textit{extrinsic}. \cite{Dyakonov71, Hirsch99, Zhang00,TsePRB05, Engel05, TsePRL06, Bruno01, Bruno05,
 Hankiewicz06, Wu08, Hankiewicz09} The spin-Hall effect observed in
Refs.\ \onlinecite{Kato04, Sih05} was shown by Engel \textit{et al.}, \cite{Engel05} Tse and Das Sarma
\cite{TsePRL06} and Hankiewicz and Vignale\cite{Hankiewicz06} to be due to extrinsic mechanisms. Band structure spin-orbit
interactions yield contributions termed \textit{intrinsic}.\cite{Murakami03, Sinova04, Inoue04, Shytov04, KrotkovPRB06} The spin-Hall effect
observed in Ref.\ \onlinecite{Wunderlich05, Brune08} is believed to be due
to intrinsic mechanisms. Extrinsic and intrinsic mechanisms were
broadly discussed in the context of the anomalous Hall effect (AHE),
which is the generation of a transverse charge and spin-polarization
current in response to an electric field in a ferromagnetic medium.
\cite{Luttinger54, Luttinger, Smit55, Smit58, Kohn57, Berger70a,
 Berger72, Nozieres72, Nozieres, Jungwirth02, Yao04, Nagaosa06,
 Borunda07, Sinitsyn07, Sinitsyn08, Nagaosa09} In fact, extrinsic
contributions to the anomalous Hall and spin-Hall effects are
closely related. \cite{TsePRL06, Engel05} The interplay of intrinsic
and extrinsic contributions is a complicated problem. It was first
addressed by Tse and Das Sarma \cite{TsePRB06} using an approach
based on the diagrammatic Kubo formula. This was followed by a
series of publications \cite{Hu06, Hankiewicz08, Raimondi09} and
this topic continues to be an active area of research. However, our
focus in this work is on the case of extrinsic spin-orbit
interactions alone and specifically the way they may be obtained
from a kinetic equation approach.

Extrinsic contributions to the spin current are of two kinds. The
first and more intuitive contribution arises from the asymmetric
scattering of up and down spins known as \textit{skew scattering}.
This effect is found beyond the first Born approximation, i.e.,
provided one goes to at least third order in the electron-impurity
potential. \cite{Kohn57, TsePRL06, Engel05} The effect appears
naturally in the standard Boltzmann collision integral
\cite{Nozieres, Hankiewicz06} provided one goes beyond the
first-Born approximation. The associated spin-Hall conductivity
scales with disorder as the ordinary Drude
conductivity \cite{TsePRL06, Engel05} (i.e., proportional to the
electron-impurity scattering time), although it is of course much
smaller.

The second extrinsic contribution has been known in the literature
as \textit{side-jump} and has two main characteristics: (i) it
appears already in the first Born approximation for the
electron-impurity potential, and (ii) the associated spin Hall
conductivity is independent of the electron-impurity scattering time
-- a surprising universality which will be fully explained below. In
contrast to skew-scattering, the side-jump mechanism cannot be
derived straightforwardly from the standard form of the Boltzmann
equation. Very early on, Luttinger \cite{Luttinger} calculated the
side jump contribution to the charge
conductivity using a recursive density matrix approach, providing a
thorough calculation yet leaving many questions unanswered
concerning the physical picture behind different contributions.
Later, Berger \cite{Berger70a, Berger72} used a wave-packet
formalism to build a semiclassical picture of side jump, identifying
it with a shift of the center of mass of the wave packet during
collisions. Nozi\`eres and Lewiner \cite{Nozieres} used this picture
and carefully studied the side jump contributions within a Boltzmann
approach. These authors accounted for six contributions, some of
which cancel each other; yet they did not associate any physical
interpretation with the cancelations. In Ref.\ \onlinecite{TsePRB05} the skew-scattering and side-jump terms were
identified also within the framework of a drift-diffusion approach. In Ref.~\onlinecite{Hankiewicz06} it was claimed that the proper way
to describe the side-jump effect is to replace the band energy in the usual
Boltzmann collision integral by a modified band energy which
includes corrections due to the spin-orbit interaction with the electric
field. However, the validity of this claim has not been formally
shown to date. Skew scattering and side jump were rigorously studied by Tse and Das Sarma \cite{TsePRL06} using a diagrammatic Kubo formalism, demonstrating that the side-jump term can be identified with an anomalous current which gives rise to a
renormalization of the current vertex. A derivation based on the Kubo formula was also presented in Ref.~\onlinecite{HankiewiczPRL06}, finding that the side-jump contribution for the conduction band is independent of disorder and of the Coulomb potential to all orders in the 
strength of these interactions. 

The purpose of this paper is to show that a rigorous alternative derivation of the side jump within the framework of the kinetic equation which does not require intuitive approaches, and follows cleanly and clearly from the Boltzmann equation provided one constructs the collision integral with care. This construction, however, requires that we go beyond the Boltzmann equation formalism and resort to the full quantum Liouville equation for the one-particle density matrix, which we treat to second order in the electron-impurity potential and to first order in the electric field. In order to focus on the essential physics, we assume that band structure spin-orbit interactions are negligible. Thus we only take into account spin-orbit interactions with the impurities and with the external electric field that drives the current.

Starting from the quantum Liouville equation we show rigorously that
the conservation of energy in the Boltzmann collision integral must
be modified to include the effective band energy, i.e., the bare
band energy plus the spin-orbit interaction with the electric field.
This modification is reflected in the appearance of a correction to
the scattering term usually found in the Born approximation. This
additional term, which is spin-dependent and linear in the electric
field, acts as a source for the spin current, yielding one half of
the side jump contribution.

The other half emerges when one takes into account the change in the
position of the particle during collisions with impurities. The
semiclassical picture of the renormalization of the trajectory of
the electron during collisions will be derived rigorously from the
scattering of electrons off the impurity potential which involves
the terms off-diagonal in the wave vector in the density matrix. It
turns out that this process and the modified conservation of energy
give equal contributions to the spin-Hall current, so that the full
side jump contribution can indeed be obtained by including twice the
spin-orbit interaction energy with the electric field in the
$\delta$-function of conservation of energy in the ordinary
Boltzmann collision integral, as suggested previously in
Ref.~\onlinecite{Hankiewicz06}. However, the present derivation is
rigorous, and constitutes formal validation of this heuristic
approach. We note that the factor of two emerges naturally from the
diagrammatic Kubo formula, as demonstrated in Ref.\
\onlinecite{TsePRL06}, when one takes into account the vertex
renormalization of the spin and charge currents.

In addition to constructing the picture of side jump outlined above,
the analysis expounded in this work supports the argument that the
full velocity operator contains the net force acting on the system,
which must vanish on physical grounds. \cite{Hankiewicz06} To
provide a formal derivation of this fact, we show that the total
contribution of the additional terms in the velocity operator to the
spin-Hall current is indeed zero. This is due to the fact that the
velocity operator contains a spin-dependent term linear in the
electric field, which reflects the spin-orbit interaction with the
electric field, as well as an additional spin-dependent term
off-diagonal in wave vector, which reflects the spin-orbit
interaction with the impurities. The importance of this latter term was recognized already in diagrammatic linear response theory in Ref.\ \onlinecite{TsePRL06}.
Within the kinetic equation formalism, the two terms in the velocity operator produce equal and
\textit{opposite} contributions to the spin-Hall current which
cancel each other out. The physical interpretation of this fact is
that, in the steady state, the average force acting on an electron
must vanish. 

In this paper, therefore, we strive to provide an understanding of
the side-jump mechanism in the absence of intrinsic spin precession
due to band-structure spin-orbit coupling, which is rigorous and at
the same time physical. We believe that a rigorous physical
understanding is a first step towards building a consistent picture,
within the kinetic equation framework, of the interplay of spin
precession due to band structure spin-orbit coupling and spin-orbit
coupling due to impurities. This interplay has been studied by Tse and Das Sarma\cite{TsePRB06} based on a diagrammatic
linear-response approach, and by Hu \textit{et al}\cite{Hu06} numerically. More recently Hankiewicz
and Vignale \cite{Hankiewicz08} constructed a phase diagram of this problem, while Raimondi and Schwab
\cite{Raimondi09} employed a Keldysh Green's function technique. The long-term aim of this work is to build a
rigorous understanding, based on the kinetic equation, of intrinsic
spin precession, skew scattering and side jump on an equal footing.

This article is organized as follows. In section II a kinetic
equation is derived starting from the quantum Liouville equation for
the density operator. In section III we explicitly construct the
collision integral, including all contributions in the first Born
approximation arising from the modification of the position
operator. Section IV discusses the contributions of the side-jump
mechanism to the spin-Hall effect. It demonstrates that the
corrections to the velocity operator do not contribute to the spin
current. We end with conclusions.

\section{Time evolution of the density matrix}

We outline in this section the formalism that will be used to
determine the collision integral and the way it gives rise to the
side jump spin-Hall current. The system is described by a density
operator $\hat{\rho}$, which obeys the quantum Liouville equation
\begin{equation}
\td{\hat{\rho}}{t} + \frac{i}{\hbar} \,
[\hat{H}_0 + \hat{H}_E + \hat{H}_U, \hat{\rho}] = 0 \,.
\end{equation}
In this equation
\begin{equation}
  \hat{H}_0 = \frac{\hbar^2 \hat{k}^2}{2m^*}
\end{equation}
is the Hamiltonian for a parabolic conduction band, with $m^*$ the
carrier effective mass, while
\begin{subequations}
  \label{eq:eham}
  \begin{eqnarray}
    \hat{H}_E & = & e \bm{E} \cdot \hat{\bm{r}} +
    \lambda \, e \bm{E} \cdot \hat{\bm{\sigma}} \times \hat{\kk} \\
    & \equiv & e \bm{E} \cdot \hat{\bm{r}}
    + {\textstyle\frac{1}{2}}\, \hat{\bm{\sigma}} \cdot \hat{\bm \Delta}_{\kk}
  \end{eqnarray}
\end{subequations}
represents the interaction with a constant uniform external electric
field $\bm{E}$. The term $\hat{H}_E$ includes both the direct
interaction and the interaction via the (material dependent)
spin-orbit coupling of strength $\lambda$. We have used the notation
$\hat{\bm \Delta}_{\kk} = 2 \lambda \, e \hat{\kk} \times \bm{E} $
for the effective field characterizing the spin-dependent part of
this interaction, a term which, as we shall see later, gives rise to
one half of the side jump contribution to the spin-Hall
conductivity. Finally
\begin{equation}
  \hat{H}_U = U (\hat{\bm{r}})+ \lambda \bm{\nabla}
  U(\hat{\bm{r}}) \cdot \hat{\bm{\sigma}} \times \hat{\kk}
\end{equation}
denotes the interaction with a set of randomly distributed
impurities, again both directly and via spin-orbit coupling.

We project the Liouville equation onto a set of time-independent
states $\{ \ket{\kk s} \}$ of definite wave vector $\kk$ and spin
orientation $s= \pm 1$ along the $z$-axis. The matrix elements of
$\hat{\rho}$ in this basis form the spin density matrix and are
written as $\rho_{\kk\kk'} \equiv \rho^{ss'}_{\kk\kk'} = \bra{\kk s}
\hat{\rho} \ket{\kk's'}$, with corresponding notations for the
matrix elements of $\hat{H}_0$, $\hat{H}_E$ and $\hat{H}_U$. We
assume that impurities are uncorrelated and the normalization is
such that the configurational average of $\bra{\kk s} \hat{H_U}
\ket{\kk's'} \bra{\kk's'} \hat{H_U} \ket{\kk s}$ is $(n_i/V) \:
|\bar{U}_{\kk\kk'}|^2 \, \delta_{ss'}$, where $n_i$ is the impurity
density, $V$ the crystal volume, and
\begin{equation}
  \label{Uso}
  \bar{U}_{\kk\kk'} = \mathcal{U}_{\kk\kk'} \,
  \big(1 - i\lambda \bm{\sigma} \cdot\kk\times\kk' \big).
\end{equation}
Here $\mathcal{U}_{\kk\kk'}$ are the matrix elements of the
electron-impurity potential $U(\hat {\bm{r}})$ between plane waves, while
$\bar{U}_{\kk\kk'}$ is reserved for the total potential of a single
impurity including the spin-orbit contribution. In what follows spin
indices will be suppressed and all quantities are assumed to be
matrices in spin space. Note that we use the convention that
$\bar{U}_{\kk\kk'}$ and $\mathcal{U}_{\kk\kk'}$ have units of energy
$\times$ volume.

The density matrix $\hat{\rho}$ is divided into a part diagonal in
$\kk$ and a part off-diagonal in $\kk$, given by $\rho_{\kk\kk'} =
f_\kk \, \delta_{\kk\kk'} + g_{\kk\kk'}$. These two parts of
$\hat{\rho}$ satisfy a set of coupled equations
\begin{subequations}
  \begin{eqnarray}
    \label{eq:FermiJfk}
    \td{f_\kk}{t} + \frac{i}{\hbar} \, [\hat{H}, \hat{f}]_{\kk\kk}
    &=& - \frac{i}{\hbar} \, [\hat{H}_U, \hat{g}]_{\kk\kk} \\ [1ex]
    \label{eq:Jfk}
    \td{g_{\kk\kk'}}{t}
    + \frac{i}{\hbar} \, [\hat{H}, \hat{g}]_{\kk\kk'}
    &=& - \frac{i}{\hbar} \, [\hat{H}_U, \hat{f}]_{\kk\kk'}
    - \frac{i}{\hbar} \, [\hat{H}_U, \hat{g}]_{\kk\kk'},
    \hspace*{2em}
\end{eqnarray}
\end{subequations}
where $\hat{H} \equiv \hat{H}_0 + \hat{H}_E$. In the first Born
approximation the solution to Eq.\ (\ref{eq:Jfk}) for $g_{\kk\kk'}$
is
\begin{equation}
\label{g}
g_{\kk\kk'} = - \frac{i}{\hbar} \lim_{\eta\rightarrow 0}
\int_0^{\infty} \!\! dt'\, e^{-\eta t'}\,  e^{- i \hat{H} t'/\hbar}
[\hat{H}_U, \hat{f} (t - t')]\, e^{i \hat{H} t'/\hbar} \big|_{\kk\kk'} ,
\end{equation}
where $\eta>0$ is a regularization factor. We are considering
variations which are slow on the scale of the momentum relaxation
time, thus we approximate $\hat{f} (t - t')$ in the integral by
$\hat{f} (t)$, \cite{Rammer} which is written simply as $\hat{f}$,
and satisfies the equation
\begin{equation}\label{kineq}
  \td{f_\kk}{t} + \frac{i}{\hbar} \, [\hat{H}, \hat{f}]_{\kk\kk}
  + \hat{J}(f_\kk) = 0,
\end{equation}
where the scattering term $\hat{J}(f_\kk)$ is
\begin{subequations}
  \label{eq:Jtgen}
  \begin{eqnarray}
    \label{eq:Jtcom}
    \hat{J}(f_\kk)
    & = & (i/\hbar) \, [\hat{H}_U, \hat{g}]_{\kk\kk} \\[2ex]
    & = & \frac{1}{\hbar^2} \lim_{\eta\rightarrow 0}
    \int_{0}^\infty \!\! dt'\, e^{- \eta t'}
    [\hat{H}_U, e^{- i \hat{H} t'/\hbar}
    [\hat{H}_U, \hat{f}]\, e^{ i \hat{H} t'/\hbar}] \big|_{\kk\kk}.
    \hspace{-2em} \nonumber \\ \label{eqJt}
  \end{eqnarray}
\end{subequations}
Equations (\ref{kineq}) and (\ref{eq:Jtgen}) describe the dynamics
of the density matrix and constitute the complete set of tools we
require in order to derive the kinetic equation satisfied by the
density matrix in an electric field, including the side-jump
contribution to the scattering term due to the modification of the
position operator by the spin-orbit interaction.

\section{Derivation of the collision integral}

We want to evaluate further the collision integral (\ref{eq:Jtgen}).
For this purpose, we decompose the matrix $f_\kk$ into a part scalar
in spin space and a spin-dependent part, thus $f_\kk = n_\kk\,
\openone + S_\kk$, with $\openone$ the identity matrix and $S_\kk$
expressible in terms of the Pauli matrices. We will show in the
following that, to first order in $\lambda$ and in the electric
field, we can write the scattering term as
\begin{equation}
  \hat{J}(f_\kk) = \hat{J}_0(n_\kk)
  + \hat{J}^a_\mathrm{sj}(n_\kk) + \hat{J}^b_\mathrm{sj}(n_\kk)
  + \hat{J}_0(S_\kk).
\end{equation}
The first of these terms comes from the band Hamiltonian
$\hat{H}_0$, is a scalar in spin space and does not depend on
$\lambda$ or on the electric field. The second term reflects the
fact that the total energy must be conserved during scattering
events, including the second term in Eq.\ (\ref{eq:eham}). The third
term comes from the direct interaction with the electric field
$e\bm{E}\cdot \hat{\bm{r}}$, and arises because $\bm{r}$ fails to
commute with the spin-orbit interaction with the impurities.
Physically this reflects the change in $\bm{r}$ during a collision
with an impurity. The resulting variation of $e\bm{E}\cdot \bm{r}$
also contributes to the overall energy balance. Both
$\hat{J}^a_\mathrm{sj}(n_\kk)$ and $\hat{J}^b_\mathrm{sj}(n_\kk)$
are spin-dependent and are linear in $\lambda$ and the electric
field $\bm{E}$. We will discuss each of the contributions in turn
below. The last scattering term, $\hat{J}_0(S_\kk)$, will be
important in the kinetic equation below in determining the
steady-state correction linear in ${\bm E}$ to the spin-dependent
part of the density matrix, and thus to the spin-Hall current.

\subsection{Scattering correction due to the conservation of the
 modified carrier energy}

In this subsection we focus on the part of the scattering term which
is linear in the electric field and arises from the addition of the
spin-orbit interaction with the electric field $\frac{1}{2}\,
\bm{\sigma}\cdot \bm{\Delta}_\kk$ to the particle energy
$\varepsilon_{0\kk} \equiv \hbar^2k^2/(2m^\ast)$. Since we are
working to first order in $\lambda$ and in the electric field, in
this subsection we only need to consider the scalar part of the
impurity potential, $\mathcal{U}_{\kk\kk'}$. Moreover the time
evolution operators in this subsection include the side jump energy
$\frac{1}{2}\, \bm{\sigma}\cdot\bm{\Delta}_\kk$, but not the term
$e\bm{E}\cdot\bm{r} $, which will be considered in the next
subsection. The matrix elements of $\hat{H}$ in the exponents of the
time evolution operators are thus diagonal in $\kk$.

Writing out all the terms in the double commutator
(\ref{eqJt}) we find
\begin{widetext}
\begin{equation}\label{Jn}
\arraycolsep 0.3ex
\begin{array}[b]{>{\displaystyle}r>{\displaystyle}l}
  \frac{1}{\hbar^2}\, [\hat{H}_U, e^{- i \hat{H} t'/\hbar}[\hat{H}_U, \hat{f}]
  \, e^{ i \hat{H} t'/\hbar}]_{\kk\kk} =
  \frac{n_i}{\hbar^2} \, \int \!\frac{d^dk'}{(2\pi)^d}
  & \big( \mathcal{U}_{\kk\kk'}
  e^{- i H_{\kk'} t'/\hbar} \mathcal{U}_{\kk'\kk} f_\kk
  e^{ i H_\kk t'/\hbar} - \mathcal{U}_{\kk\kk'}
  e^{- i H_{\kk'} t'/\hbar} f_{\kk'} \mathcal{U}_{\kk'\kk} \,
  e^{ i H_\kk t'/\hbar}
  \\ [3ex]
  & - e^{- i H_\kk t'/\hbar} \mathcal{U}_{\kk\kk'} f_{\kk'}
  e^{ i H_{\kk'} t'/\hbar} \mathcal{U}_{\kk'\kk}
  + e^{- i H_\kk t'/\hbar} f_\kk \mathcal{U}_{\kk\kk'}
  e^{ i H_{\kk'} t'/\hbar} \mathcal{U}_{\kk'\kk} \big)\, ,
\end{array}
\end{equation}
where $H_\kk = \varepsilon_{0\kk} + \frac{1}{2}\,
\bm{\sigma}\cdot\bm{\Delta}_\kk$ and $d$ is the dimensionality of
the system. By expanding the exponentials of Pauli matrices, the
product of time evolution operators can be written, to first order
in $\lambda$, as
\begin{equation}
e^{- i H_{\kk'} t'/\hbar} e^{ i H_\kk t'/\hbar} =
e^{ i(\varepsilon_{0\kk} - \varepsilon_{0\kk'}) t'/\hbar}
\left[\cos \frac{\Delta_\kk t'}{2\hbar} \cos\frac{\Delta'_\kk t'}{2\hbar}
- i\, \bm{\sigma}\cdot\hat{\bm{\Delta}}_{\kk'}\,
\cos\frac{\Delta_\kk t'}{2\hbar} \sin\frac{\Delta'_\kk t'}{2\hbar}
+ i\, \bm{\sigma}\cdot\hat{\bm{\Delta}}_\kk\,
\sin\frac{\Delta_\kk t'}{2\hbar}\cos\frac{\Delta'_\kk t'}{2\hbar}\right]\, ,
\end{equation}
where $\hat{\bm{\Delta}}_\kk$ is a unit vector in $\bm{\Delta}_\kk$
direction.

The only task that remains is integration over the time variable
$t'$, giving a series of $\delta$-functions reflecting energy
conservation. The overall result for this scattering term, to first
order in $\lambda$, can be decomposed into a scalar part $\hat{J}_0
\, (n_\kk)$ independent of $\lambda$, and spin-dependent parts
$\hat{J}^a_\mathrm{sj} \, (n_\kk) + \hat{J}_0\, (S_\kk)$. These
parts may be written as follows
\begin{subequations}
\begin{eqnarray}
\hat{J}_0 \, (n_\kk) & = & \frac{\pi n_i}{2\hbar}\, \int\!\frac{d^dk'}{(2\pi)^d}
\: |\mathcal{U}_{\kk\kk'}|^2 (n_\kk - n_{\kk'})
\big[ \delta(\epsilon_+ - \epsilon'_+) + \delta(\epsilon_- - \epsilon'_-)
+ \delta(\epsilon_+ - \epsilon'_-) + \delta(\epsilon_- - \epsilon'_+) \big]
\label{J0} \\ [2ex]
\hat{J}^a_\mathrm{sj} \, (n_\kk) & = & \frac{\pi n_i}{2\hbar}\,
\int\!\frac{d^dk'}{(2\pi)^d} \: |\mathcal{U}_{\kk\kk'}|^2(n_\kk - n_{\kk'})\{
\bm{\sigma}\cdot(\hat{\bm{\Delta}}_\kk + \hat{\bm{\Delta}}_{\kk'})\,
[\delta(\epsilon_+ - \epsilon'_+) - \delta(\epsilon_- - \epsilon'_-)]
\nonumber \\ & & \hspace*{12em}
+ \bm{\sigma}\cdot(\hat{\bm{\Delta}}_\kk
- \hat{\bm{\Delta}}_{\kk'})\, [\delta(\epsilon_+ - \epsilon'_-)
- \delta(\epsilon_- - \epsilon'_+)] \} \\ [2ex]
\hat{J}_0 \, (S_\kk) & = & \frac{2\pi n_i}{\hbar}\, \int\!\frac{d^dk'}{(2\pi)^d}
\: |\mathcal{U}_{\kk\kk'}|^2 (S_\kk - S_{\kk'})
 \delta(\varepsilon_{0\kk} - \varepsilon_{0\kk '}).
\end{eqnarray}
\end{subequations}
The full energies $\epsilon_\pm = \varepsilon_{0\kk} \pm
\Delta_\kk/2$ and $\epsilon'_\pm = \varepsilon_{0\kk'} \pm
\Delta_{\kk'}/2$, where $\Delta_\kk = |\bm{\Delta}_\kk|$. The scalar
term $\hat{J}_0\, (n_\kk)$ reproduces the ordinary
Boltzmann-equation scattering term.
The side-jump term $\hat{J}^a_\mathrm{sj}\, (n_\kk)$ constitutes a
correction that reflects the presence of the spin-orbit interaction
energy with the electric field in the condition for energy
conservation. We expand the $\delta$-functions in this scattering
term in $\Delta_\kk$ as
\begin{equation}
\delta(\epsilon_+ - \epsilon'_+) = \delta(\varepsilon_{0\kk} -
\varepsilon_{0\kk'}) + \bigg(\frac{\Delta_\kk}{2} -
\frac{\Delta_{\kk'}}{2}\bigg) \, \pd{}{\varepsilon_{0\kk}}\,
\delta(\varepsilon_{0\kk} - \varepsilon_{0\kk'}),
\end{equation}
with corresponding expressions for the other combinations of
$\delta$-functions. Adding all contributions together the scattering
term $\hat{J}^a_\mathrm{sj}\, (n_\kk)$ simplifies considerably and
we obtain the final expression
\begin{equation}\label{Jsjn}
\hat{J}^a_\mathrm{sj} \, (n_\kk) = \frac{2\pi n_i}{\hbar}\,
\int\!\frac{d^dk'}{(2\pi)^d} \: |\mathcal{U}_{\kk\kk'}|^2(n_\kk - n_{\kk'})\,
\frac{1}{2} \, \bm{\sigma}\cdot(\bm{\Delta}_\kk - \bm{\Delta}_{\kk'}) \,
\pd{}{\varepsilon_{0\kk}} \, \delta(\varepsilon_{0\kk} - \varepsilon_{0\kk'}).
\end{equation}
\end{widetext}
The presence of this scattering term reflects the fact that the
total energy including the spin-orbit interaction energy with the
electric field is conserved in elastic collisions.

\subsection{Scattering correction due to the change in $\bm{r}$
 during collisions}

We have so far ignored the presence of the term
$e\bm{E}\cdot\hat{\bm{r}}$ in the time evolution operator. At this
stage we would like to determine the additional scattering term
linear in $\bm{E}$ arising from it, which we denote by
$\hat{J}^b_\mathrm{sj} \, (n_\kk)$. To accomplish this we use Eq.\
(\ref{g}) to find the correction $g^b_{\kk\kk'}$ to $g_{\kk\kk'}$
arising from the presence of $e\bm{E}\cdot\hat{\bm{r}}$ in the time
evolution operator. Using Eq.\ (\ref{eq:Jtcom}) we will then obtain
$\hat{J}^b_\mathrm{sj} \, (n_\kk)$ as $(i/\hbar) \, [\hat{H}_U, \hat
g^b]_{\kk\kk}$.

Starting from Eq.\ (\ref{g}), we expand the time evolution operator
to first order in the term $e\bm{E}\cdot\hat{\bm{r}}$. Using the
matrix elements of the ordinary position operator $\hat{\bm{r}}$
between Bloch states
\begin{equation}
\bra{\kk} \hat{\bm{r}} \ket{\kk'} = i \, \pd{}{\kk}\, \delta(\kk - \kk'),
\end{equation}
we obtain additional terms of the form
\begin{equation}
t' e^{- i \varepsilon_{0\kk} t'} e\bm{E}\cdot\pd{}{\kk}\delta(\kk - \kk')
= \bigg(\frac{i\,\partial}{\partial\varepsilon_{0\kk}}
e^{- i \varepsilon_{0\kk} t'} \bigg)\, e\bm{E}\cdot\pd{}{\kk}\delta(\kk - \kk').
\end{equation}
We integrate over $t'$ as before, and after some lengthy but
straightforward algebra, we obtain
\begin{widetext}
\begin{equation}\label{gb}
g^b_{\kk\kk'} = - \pi e\bm{E} \cdot \int\frac{d^dk'}{(2\pi)^d} \:
\bigg(\pd{\bar{U}_{\kk\kk'}}{\kk} + \pd{\bar{U}_{\kk\kk'}}{\kk'} \bigg)
\, (n_\kk - n_{\kk'}) \, \pd{}{\varepsilon_{0\kk}} \,
\delta(\varepsilon_{0\kk} - \varepsilon_{0\kk'}).
\end{equation}
We have not written out explicitly a contribution to $g^b_{\kk\kk'}$
containing terms of the form $\partial n_\kk/\partial\kk$. We find
that such terms drop out in the final evaluation of spin currents
when the scattering potential is elastic, as a result of integrating
over $\kk$ and $\kk'$. In the final analysis these terms involve the
square matrix element $|\bar{U}_{{\bm k}{\bm k}'}|^2$ which does not
have any contributions linear in $\lambda$. We find that the leading
contribution due to these terms is $\propto \lambda^2$ and may
therefore be neglected. Evaluating the $\kk$-derivatives of the
impurity potentials gives
\begin{equation}
\pd{\bar{U}_{\kk\kk'}}{\kk} + \pd{\bar{U}_{\kk\kk'}}{\kk'} =
- i\lambda \mathcal{U}_{\kk\kk'} \, \bm{\sigma}\times\big(\kk - \kk' \big).
\end{equation}
Substituting this into the Eq.\ (\ref{gb}) and subsequently
evaluating $\hat{J}^b_\mathrm{sj} \, (n_\kk) = (i/\hbar) \, [\hat{H}_U,
\hat{g}^b]_{\kk\kk}$ we obtain the scattering term
\begin{equation}
\hat{J}^b_\mathrm{sj} \, (n_\kk) = \frac{2\pi n_i e \lambda}{\hbar}
\int\frac{d^dk'}{(2\pi)^d} \: |\mathcal{U}_{\kk'\kk}|^2  \, (n_\kk - n_{\kk'})
\, \bm{E} \cdot\bm{\sigma} \times\big(\kk - \kk' \big)
\pd{}{\varepsilon_{0\kk}} \, \delta(\varepsilon_{0\kk} - \varepsilon_{0\kk'}),
\end{equation}
\end{widetext}
which is easily seen to be exactly equal to $\hat{J}^a_\mathrm{sj}
(n_\kk) $. The sum of these terms constitutes the total
\textit{side-jump scattering term} $\hat{J}_\mathrm{sj}(n_\kk) =
\hat{J}^a_\mathrm{sj} (n_\kk) + \hat{J}^b_\mathrm{sj}(n_\kk) = 2
\hat{J}^a_\mathrm{sj} (n_\kk) $, which contains the well-known
factor of two associated with side jump. \cite{Nozieres} We
emphasize that we obtain this reinforcement of the side jump
directly from the scattering term, and our work shows no evidence
that it is related in any direct way to the integral of the
velocity operator [see Eqs.~(\ref{diagvel}) and (\ref{offdiagvel})
below] over the time of a collision. \cite{Engel05}

\section{Contribution of the side-jump mechanism to the spin-Hall
 current}

We have derived a contribution linear in the electric field to the
scattering term appearing in the kinetic equation. This contribution
is brought about by the spin-dependent interaction of the charge
carriers with the electric field due to the spin-orbit interaction.
In this section we will first evaluate the correction that this term
yields in the spin-dependent part of the density matrix, and we will
show that this correction accounts fully for the side-jump spin-Hall
current including the important factor of two. \cite{Nozieres, Engel05, TsePRL06} This
is done in subsection \ref{sec:SJSHE}. To show that this is the only
side-jump contribution to the spin-Hall effect, subsection
\ref{sec:velop} will demonstrate that the modifications to the
velocity operator do not contribute any additional terms to the spin
current.

\subsection{Contribution of the side-jump scattering term}
\label{sec:SJSHE}

We need to find the contribution to the spin-Hall current brought
about by the additional scattering term $\hat{J}_\mathrm{sj}
(n_\kk)$. In order to further evaluate Eq.\ (\ref{kineq}) we let
$f_\kk = f_{0\kk} + f_{E\kk}$, where $f_{0\kk} (\varepsilon_{0\kk})$
is the Fermi-Dirac function, which is a scalar in spin space in the
case under study, and $f_{E\kk}$ is the correction we will determine
from the kinetic equation. Firstly, the kinetic energy part of the
Hamiltonian, $\varepsilon_{0\kk}$, drops out of the commutator.
Also, in the commutator $[\frac{1}{2} \, \bm{\sigma}
\cdot\bm{\Delta}_\kk, f_\kk]$ we note that $\frac{1}{2} \,
\bm{\sigma}\cdot\bm{\Delta}_\kk$ is first order in the electric
field, so the density matrix can be replaced with the equilibrium
Fermi-Dirac function, which is a scalar in spin space, so
$[\frac{1}{2} \, \bm{\sigma}\cdot\bm{\Delta}_\kk, f_{0\kk}] = 0$.
Moreover, in the side-jump scattering term $\hat{J}_\mathrm{sj}
(n_\kk)$, which is also first-order in the electric field, we may
replace $n_\kk$ by $f_{0\kk}$. Following some short and
straightforward algebra, the side-jump scattering term can be
written as
\begin{equation}
\hat{J}_\mathrm{sj} (f_{0\kk}) = \frac{1}{\tau_p} \,
\bm{\sigma}\cdot\bm{\Delta}_\kk \, \delta(\varepsilon - \varepsilon_F),
\end{equation}
where $\tau_p$ is the usual momentum relaxation time.

Next, we decompose $f_{E\kk}$ into a part scalar in spin space and a
spin-dependent part, thus $f_{E\kk} = n_{E\kk}\, \openone +
S_{E\kk}$. The equation for $n_{E\kk}$ in the steady state is the
ordinary scalar Boltzmann equation. The term $[\hat{H}_E,
\hat{f}]_\kk$ becomes $(e/\hbar) \bm{E} \cdot (\partial f_0
/ \partial \kk)$ which is a scalar and acts as the source term for
$n_{E\kk}$. We write the scalar part of the Boltzmann equation as
\begin{equation}\label{Boltzn}
\hat{J}_0(n_{E\kk}) = \frac{e\bm{E}}{\hbar}\cdot \pd{f_{0\kk}}{\kk},
\end{equation}
where the scattering term $\hat{J}_0(n_{\kk})$ has been defined in
Eq.\ (\ref{J0}). The solution for $n_{E\kk}$ is written as
\begin{equation}\label{n}
n_{E\kk} = \frac{\tau_p \, e\bm{E}}{\hbar}\cdot \pd{f_{0\kk}}{\kk}
= \frac{\hbar\tau_p \, e\bm{E}\cdot\kk}{m^\ast}
\pd{f_{0\kk}}{\varepsilon_{0\kk}}.
\end{equation}
The equation for $S_{E\kk}$ in the steady state is $\hat{J}_0
(S_{E\kk}) = - \hat{J}_\mathrm{sj} (f_{0\kk}) $, in which the RHS,
$\hat{J}_\mathrm{sj} (f_{0\kk})$, acts as a source term for
$S_{E\kk}$. Substituting the explicit expressions for the two
scattering terms, this equation can be written in a simpler form as
\begin{equation}
\frac{S_{E\kk}}{\tau_p} = - \frac{1}{\tau_p}\,
\bm{\sigma}\cdot\bm{\Delta}_\kk \, \delta(\varepsilon - \varepsilon_F),
\end{equation}
and has the simple solution
\begin{equation}
S_{E\kk} = - \bm{\sigma}\cdot\bm{\Delta}_\kk \,
\delta(\varepsilon - \varepsilon_F).
\end{equation}
We would like to draw attention to the fact that the additional
side-jump collision integral is part of the source for $S_{E\kk}$.
The source term itself contains a factor of $1/\tau_p$, which
cancels the $1/\tau_p$ appearing on the LHS of the equation for
$S_{E\kk}$. This explains why the end result for the side jump
contribution to the spin-Hall current does not depend on the
strength and shape of the impurity potential.

Now that we have found the solution for $S_{E\kk}$, in other words
the spin-dependent part of the density matrix in an electric field,
we can determine its contribution to the spin-Hall current. We
denote the components of the spin velocity as $v^i_j$ which
corresponds to a spin component $i$ flowing along the direction $j$.
In finding the contribution to the spin-Hall current due to
$S_{E\kk}$ we can restrict ourselves to the term in the spin
velocity to zeroth order in the electric field, which is $v^i_j =
(\hbar k_j / m^\ast) \, \hbar\sigma_i/ 2$. For $\bm{E} =
(E_x,0,0)$ the side-jump Hamiltonian $\frac{1}{2} \,
\bm{\sigma}\cdot\bm{\Delta}_\kk$ becomes
\begin{equation}
e \lambda \bm{E}\cdot \bm{\sigma} \times \kk = - e\lambda E_x k_y \sigma_z,
\end{equation}
This gives a side-jump spin-Hall current as a result of the
modification to the scattering term
\begin{equation}\label{jsct}
  j^z_y\Big|^\mathrm{sct} = \bigg(\frac{\hbar}{2}\bigg)
  \int\frac{d^dk}{(2\pi)^d} \: \frac{\hbar k_y}{m^\ast} \,
  \trace (\sigma_z S_{E\kk}) = ne\lambda \, E_x,
\end{equation}
where the trace is taken over the spin components and $n$ is the
density. This term therefore gives a spin-Hall conductivity
$\sigma^z_{yx}|^\mathrm{sct} = ne\lambda$, which is the usual
side-jump term in the spin-Hall current. \cite{Engel05, TsePRL06, TsePRB06,
 Nozieres, Hankiewicz06, Hankiewicz09} This result is valid in both
two and three dimensions.

\subsection{Vanishing contribution of the corrections to the
 velocity operator}
\label{sec:velop}

It will be shown in this subsection that the correction to the velocity
operator linear in the electric field does not yield any additional
terms in the spin-Hall current. The velocity operator is defined as
the time derivative of the physical position operator, which in turn
is given by
\begin{equation}
\hat{\bm{r}}_\mathrm{phys}=\hat{\bm{r}}+\lambda
\hat{\bm{\sigma}} \times \hat{\kk}\, .
\end{equation}
Notice that all spin-orbit interactions can be most directly derived
by replacing $\bm{r}$ by $\bm{r}_\mathrm{phys}$ in the direct
interactions and expanding to first order in $\lambda$.

The velocity operator has a part which is diagonal in $\kk$
and is given by
\begin{equation}\label{diagvel}
\arraycolsep 0.3ex \hspace{0.5em}
\begin{array}[b]{r>{\displaystyle}l}
\bm{v}_\kk = & (i/\hbar) \, [\hat{H},
\hat{\bm{r}}_\mathrm{phys}]_\kk \\[1.5ex]
= & \frac{\hbar \kk}{m^\ast} + \frac{i}{\hbar}\,
[e \lambda \bm{E}\cdot \hat{\bm{\sigma}} \times \hat{\kk},  \hat{\bm{r}}]_\kk
+ \frac{i}{\hbar}\, [e \bm{E}\cdot\hat{\bm{r}},  \lambda \hat{\bm{\sigma}}
\times \hat{\kk}]_\kk\\ [3ex]
 = & \frac{\hbar \kk}{m^\ast} - \frac{2e \lambda}{\hbar}
 \, \bm{\sigma} \times \bm{E}.
\end{array} \hspace{-2.5em}
\end{equation}
The $\kk$-diagonal part of the spin velocity $v^i_j$, up to first
order in the electric field, is thus
\begin{equation}
v^i_j = \frac{\hbar k_j}{m^\ast} \, \frac{\hbar\,\sigma_i}{2}
- \frac{e \lambda}{2} \, \{ (\bm{\sigma} \times \bm{E})_j, \sigma_i \}
\end{equation}
where $\{A,B\} \equiv AB + BA$. For an electric field along
$\hat{\bm{x}}$ we obtain for the $\kk$-diagonal part of the spin
velocity $v^z_y$ the expression
\begin{equation}
v^z_y = \frac{\hbar}{2}\, \frac{\hbar k_y}{m^\ast} \,
\sigma_z - e \lambda \, E_x.
\end{equation}
The $\bm{E}$-dependent part of the spin velocity operator is a
scalar in spin space, and its contribution to the spin-Hall current
is found by multiplying by the scalar part of the equilibrium
density matrix $f_{0\kk}$. It gives us the term
\begin{equation}\label{jvelop}
j^z_y \Big|^{vel, d} = - e \lambda \, E_x \int \frac{d^d k}{(2\pi)^d}\:
\trace \, f_{0\kk} = - n e \lambda \, E_x
\end{equation}
so its contribution to the spin-Hall conductivity is
$\sigma^z_{yx} |^{vel, d} = -ne\lambda$.

This is, however, not the full story. The velocity operator also has
a term that is off-diagonal in the wave vector, which is referred to
as $\bm{v}_{\kk\kk'}$ and is given by
\begin{equation}\label{vel}
\bm{v}_{\kk\kk'} = \frac{i}{\hbar} \,
[\hat{H}_U, \hat{\bm{r}}_\mathrm{phys}]_{\kk\kk'}.
\end{equation}
The matrix elements of the impurity potential are given by Eq.\
(\ref{Uso}). The part of the matrix element $\bm{v}_{\kk\kk'}$
originating from $\hat{\bm{r}}$ is easily seen to be
\begin{equation}\label{com}
\frac{i}{\hbar} \, [\hat{U}, \hat{\bm{r}}]_{\kk\kk'} = \frac{1}{\hbar} \,
\bigg( \pd{U_{\kk\kk'}}{\kk'} + \pd{U_{\kk\kk'}}{\kk}  \bigg).
\end{equation}
The disorder potential is the potential $\hat{U}$ due to the full
ensemble of impurities present in the system. In the final result
for the spin-Hall current, the $\kk$-off-diagonal part of the
velocity operator will be traced with the $\kk$-off-diagonal part of
the density matrix $g_{\kk\kk'}$, which in the first Born
approximation is also linear in $\hat{U}$. Once this is done, a
configurational average will be performed over the impurities. In
the end we seek the result to first order in $\lambda$. However, it
proves more straightforward to work in terms of the full potential
$\hat{U}$ until the end. Only then we will restrict ourselves to the
terms which are first order in $\lambda$.

With these insights in mind we proceed, Eq.\ (\ref{vel}) yields
\begin{equation}\label{offdiagvel}
\bm{v}_{\kk\kk'} = - \frac{2i\lambda}{\hbar} \, \bm{\sigma}\times (\kk - \kk') \, U_{\kk\kk'},
\end{equation}
where we have written the matrix elements of the full potential
$\hat{U}$. Note that this expression has \textit{not} been averaged
over impurity configurations. This expression for the ${\bm
 k}$-off-diagonal part of the velocity operator holds because, for
elastic scattering, the scalar part of the scattering potential
$\mathcal{U}_{\kk\kk'} \equiv \mathcal{U}(\kk - \kk')$ depends only
on the difference $\kk - \kk'$. Its contribution to the velocity
operator is immediately seen to be zero. This can also be understood
by noting that the scalar part of the impurity potential commutes
with ${\bm r}$ and does not contribute to the velocity.

The $\kk$-off-diagonal part of the velocity operator contributes to
the side-jump spin-Hall current. To find its contribution we return
to Eq.\ (\ref{g}) and integrate over time to find
\begin{equation}\label{gprelim}
\displaystyle g_{\kk\kk'} = i \pi \, \delta(\varepsilon_{0\kk}
- \varepsilon_{0\kk'}) U_{\kk\kk'} (f_\kk - f_{\kk'}).
\end{equation}
This expression also contains the matrix elements of the full
impurity potential and has \textit{not} been averaged over impurity
configurations. 

The contribution of the $\kk$-off-diagonal part of the velocity
operator to the spin current is found by taking the trace of the
spin velocity arising from Eq.\ (\ref{offdiagvel}) with the
$\kk$-off-diagonal part of the density matrix given in Eq.\
(\ref{gprelim}). This yields for the spin-Hall current
\begin{widetext}
\begin{subequations}
  \label{jvelod}
  \begin{eqnarray}
    j^z_y\Big|^{vel, od} & = & \frac{\hbar}{2} \, \trace \sigma_z  
    \int \frac{d^dk}{(2\pi)^d} \int \frac{d^dk'}{(2\pi)^d} v^y_{\kk\kk'} 
    g_{\kk'\kk} \\[2ex] & = & - (\lambda \hbar) \frac{2\pi}{\hbar}  
    \int \frac{d^dk}{(2\pi)^d}  (k_x - k_x') \, 
    \bkt{U_{\kk\kk'} U_{\kk'\kk}} \, 
    \delta(\varepsilon_{0\kk'} - \varepsilon_{0\kk}) (f_{\kk} - f_{\kk'}),
  \end{eqnarray}
\end{subequations}
where the bracket denotes the average over impurity configurations.
At this stage we introduce the simplification that we require only
terms to first order in $\lambda$. We note that, since the entire
term in Eq.\ (\ref{jvelod}) already contains $\lambda$, the other
terms in this equation are needed only to zeroth order in $\lambda$.
Consider first the term proportional to $k_x$, which, to first order
in $\lambda$, can be written as
\begin{subequations}
\begin{eqnarray}
  - (\lambda \hbar)  \int \frac{d^dk}{(2\pi)^d}  k_x \bigg[\frac{2\pi}{\hbar}
  \int \frac{d^dk'}{(2\pi)^d}  \, \bkt{U_{\kk\kk'} U_{\kk'\kk}} \,
  \delta(\varepsilon_{0\kk'} - \varepsilon_{0\kk}) (n_{\kk} - n_{\kk'})\bigg]
   & = & - (\lambda \hbar)  \int \frac{d^dk}{(2\pi)^d} k_x
  \hat{J}_0(n_{\bm k}) \\
  & = & - (\lambda \hbar)  \int \frac{d^dk}{(2\pi)^d}  k_x
  \bigg(\frac{eE_x}{\hbar}\pd{f_{0{\bm k}}}{k_x} \bigg).
\end{eqnarray}
\end{subequations}
\end{widetext}
where the last replacement follows from the scalar Boltzmann
equation, as written in Eq.\ (\ref{Boltzn}), assuming, as before,
that ${\bm E} \parallel \hat{\bm x}$. Further, Eq.\ (\ref{jvelod})
also contains a term proportional to $k_x'$, which is easily seen to
give exactly the same contribution if one swaps ${\bm k}$ and ${\bm
 k}'$ in the summation. The contribution of the $\kk$-off-diagonal
part of the velocity operator to the spin Hall current to first
order in $\lambda$ is therefore
\begin{equation}
  j^z_y\Big|^{vel, od} = - 2\lambda eE_x \int\frac{d^dk}{(2\pi)^d} \, 
  k_x \, \pd{f_{0{\bm k}}}{k_x} = ne\lambda\, E_x.
\end{equation}
The spin-Hall conductivity originating from this term is
$\sigma^z_{yx} |^{vel, od} = ne\lambda$ and it exactly cancels the
contribution $\sigma^z_{yx} |^{vel, d}$ from the $\kk$-diagonal
$\bm{E}$-dependent part of the velocity operator. The physical
explanation of this cancellation is the following. Notice that the
full spin-dependent part of the velocity operator, from Eqs.\
(\ref{diagvel}) and (\ref{offdiagvel}) is
\begin{equation}
 - \frac{2e \lambda}{\hbar} \, \bm{\sigma} \times \bm{E}
 - \frac{2i\lambda}{\hbar} \, \bm{\sigma}\times (\kk - \kk') \,
 \mathcal{U}_{\kk\kk'}
\end{equation}
which contains the total force acting on the system. According to
Ehrenfest's theorem, the expectation values of position and momentum
obey time evolution equations analogous to those of classical
mechanics. Consequently the expectation value of the force should be
zero in the steady state, consistent with the earlier suggestion
that the total force acting on the system does not contribute to the
spin current. \cite{Hankiewicz06} We note also that the presence of the velocity terms off-diagonal in wave vector is crucial
in obtaining the correct side-jump contribution in the diagrammatic Kubo-formula approach, as demonstrated in Ref.\ \onlinecite{TsePRL06}.

\section{Summary and Discussion}

We have completed the formal derivation of the side-jump spin-Hall
current, where we have considered (in the absence of intrinsic spin
precession) all contributions to the kinetic equation in the first
Born approximation. We will now discuss our findings and their
implications. It is evident from our analysis that, within the
kinetic equation framework, the side-jump
spin-Hall current originates solely from the modification of the Boltzmann
collision integral due to the spin-dependent interaction energy of
an electron with the external electric field and the impurity field.
The density-matrix formulation of the problem shows that the
spin-orbit interaction with the electric field alters the condition
for energy conservation, since the total energy conserved during
collisions must include the spin-dependent part. In addition, the
spin-orbit coupling with the impurities causes a change in the
position of the electron during scattering process, which again
affects the energy balance via the interaction energy $e \bm{E}\cdot
\bm{r}$. This effect doubles the size of the side-jump current.

The understanding of the side-jump effect emerging from this
derivation differs from the conventional explanation,
according to which this phenomenon is attributed to the
linear-in-$\bm{E}$ modification of the velocity operator. It is
clearly seen in the previous section that the full velocity operator
in the presence of spin-orbit interactions contains an extra term
due to the impurity potential, which is off-diagonal in wave vector.
This term, and the $\kk$-diagonal velocity operator result in two
corrections to the spin current that are equal in magnitude but
opposite sign so they cancel out. This could be justified
informally using the fact that the velocity operator contains the
total force acting on the system (or rather the term to leading
order in $\lambda$ of this force), and therefore should vanish in
the steady state \cite{Hankiewicz06}. Therefore, contrary to
conventional assumptions, the side-jump contribution to the
spin-Hall current is traced to the qualitatively different carrier
spin dynamics during collisions. It is not traced to any
linear-in-$\bm{E}$ correction to the velocity operator. The linear-in-$\bm{E}$ correction to the velocity operator
is canceled by the contribution of the off-diagonal in $\kk$ correction to the velocity operator.
The importance of this off-diagonal-in-$\kk$ correction in the velocity operator in obtaining the correct side-jump current was recognized in the diagrammatic Kubo-formula approach as shown in Ref.\ \onlinecite{TsePRL06}. 

We emphasize that in obtaining our results we have used a rigorous
quantum mechanical formulation, starting with the quantum Liouville
equation and making, in the course of the derivation, the same
assumptions that are characteristically made in linear response
theory. Therefore our formalism could in some sense be regarded as
being built from the ground up. The approach we have used
demonstrates that the derivation of the side jump does not need to
rely on intuitive semiclassical ideas as long as the collision
integral is derived rigorously from the fundamental starting point
of all transport theories. In contrast, Ref. \onlinecite{Nozieres} using semiclassical Boltzmann arguments counted six
terms contributing to side jump, but did not clearly indicate
which terms should cancel. Since for the conduction band the
side-jump contributions have the same magnitude but opposite signs,
it led to freedom in choosing which terms cancel and still obtaining
the correct amplitude of the side-jump. For example Ref. \onlinecite{Engel05} counted the terms from the anomalous velocity and the
shift of the position operator. Our analysis shows exactly which
terms are non-zero and which contributions cancel and is in
agreement with the Kubo derivation of side jump contributions
presented in Refs.~\onlinecite{HankiewiczPRL06, TsePRL06}. Our results are also
in agreement with recent extensive studies of the anomalous Hall
effect in magnetic semiconductors. \cite{Sinitsyn07, Kovalev09,
 Nagaosa09} In particular, Refs.\ \onlinecite{Sinitsyn07} and \onlinecite{Sinitsyn05}, by comparing the semiclassical description of the
side-jump with the Kubo and Keldysh formalisms, found that the
velocity operator is unchanged and that the two contributions to the
side jump have different origins due to the renormalization of the
distribution function and due to the modification of the
conservation of energy.

We would like to acknowledge stimulating discussions with S.~Das
Sarma, Jairo Sinova, Hans-Andreas Engel and Peter Schwab. DC was
supported by LPS-NSA-CMTC. E. M. H. was financially supported by DFG
grant HA 5893/1-1. Work at Argonne was supported by DOE BES under
Contract No.\ DE-AC02-06CH11357. GV acknowledges support from NSF
Grant No. DMR-0705460.

\bibliography{review}

\end{document}